\documentclass[12pt]{article}
\usepackage{graphicx}
\usepackage{amsmath}
\usepackage{amssymb}

\begin{document}

\begin{titlepage}
\begin{flushright}
OSU-HEP-01-07\\
\end{flushright} \vskip 2cm
\begin{center}
{\Large\bf Collider Implications of Kaluza-Klein \\
Excitations of the Electroweak Gauge Bosons} \vskip 1cm
{\normalsize\bf
C.D.\ McMullen\footnote{email: mcmulle@okstate.edu} and S.\
Nandi\footnote{email:
shaown@osuunx.ucc.okstate.edu}\addtocounter{footnote}{-2}
\\} \vskip 0.5cm
{\it Department of Physics, Oklahoma State University\\
Stillwater, OK~~74078, USA\\ [0.1truecm] }

\end{center}
\vskip 2.5cm

\begin{abstract}

It is possible for the SM fields to propagate into one or more
large extra compact dimensions.  Such fields have associated KK
excitations that produce additional contributions to the SM
processes.  We calculate the effects that these KK excitations
have on cross sections for $e^{+}e^{-}$ collider processes in a
model in which the SM gauge bosons, and perhaps the Higgs, can
propagate into one TeV$^{-1}$-size extra compact dimension.
Specifically, we consider muon pair production, Bhabha scattering,
dijet production, Higgs production, and single-photon production.
At prospective high energy $e^{+}e^{-}$ colliders, we find
significant deviations from the SM cross sections: For example, a
$600$ GeV collider produces a $20\%$ reduction in the cross
section for muon pair production as compared to the SM for a
compactification scale of $3.5$ TeV. The effect is much smaller at
LEP$2$ energies, where the compactification scale must be lower
than $2$ TeV in order to produce a $6\%$ effect.

\end{abstract}

\end{titlepage}

\newpage

\noindent {\bf 1.  Introduction}

\vspace{0.2cm}

\noindent There has been a flurry of recent interest in the
low-energy phenomenology of string-inspired models where the
presence of large extra compact dimensions produces a string scale
that is quite small compared to the usual four-dimensional Planck
scale~\cite{superstring}.  In this class of models based on the
approach of Arkani-Hamed, Dimopoulos, and Dvali
(ADD)~\cite{Planck}, $n$ large extra dimensions are compactified
at the same scale $R^{-1}$, where the size $R$ is related to the
four-dimensional Planck scale $M_P$ via the relation

\vspace{-8pt} \begin{equation} \label{eq:Planck} M_{P}^2 =
M_{\star}^{n+2} \, R^n \, .
\end{equation}

\noindent This new scale $M_{\star}$ is the fundamental ($4 \! +
\! n$)-dimensional Planck scale, which is of the same order as the
string scale.  It has been demonstrated that this relation
(\ref{eq:Planck}) is phenomenologically viable~\cite{Planck} for
$n \geq 2$, that $R$ can be as large as a sub-millimeter, and that
the string scale could be as small as a few tens of TeV. In the
case where all six extra dimensions from the superstring theory
are compactified at the same scale, then the compactification
scale $1/R$ is about $10$ MeV.\footnote{The bound is even lower if
the fields propagate into fewer extra
dimensions.}\addtocounter{footnote}{-1} Therefore, any Standard
Model (SM) fields that propagate into these extra dimensions (the
bulk) have Kaluza-Klein (KK) excitations with masses at the $10$
MeV scale.  The present high-energy collider limit of about a TeV
for such states thus implies, in these scenarios, that all SM
fields are confined to a three-dimensional wall (D$_3$ brane) of
the usual three spatial dimensions.  This is characteristic of the
class of models in which the compactification is symmetric,
\textit{i.e.}, all of the extra dimensions have the same
compactification radius $R$.

By contrast, in an asymmetric scenario, SM fields can propagate
into a single extra dimension with a compactification radius as
large as a TeV$^{-1}$, for example, and thus have lowest-lying KK
excitations at the TeV scale.  Aside from the obvious collider
implications that are just at the edge of the grasp of present
high-energy colliders, the phenomenology of such a model includes
an alteration in the evolution of the gauge couplings from the
usual logarithmic to power law behavior~\cite{unify}.  The result
is a unification scale that can be much smaller~\cite{unify},
perhaps as little as a few TeV.  The simplest asymmetrical
compactification scenario involves two distinct compactification
scales.  The recently analyzed case~\cite{asym} of $n$ extra
dimensions of size $R \sim$ mm and $m$ of size $r \sim$ TeV$^{-1}$
was proposed as a unique model in which both the collider and
gravitational~\cite{gravity} bounds are just beyond the reach of
present experiments.  In particular, the $n=1, \, m=5$ case was
shown to satisfy all of the current astrophysical and cosmological
constraints~\cite{astro}.  The scaling relation for this model
is~\cite{asym}

\vspace{-8pt} \begin{equation} M_{P}^2   =   M^8 \, R \, r^5 \, ,
\end{equation}

\noindent where $M$ is the unification scale.  As a numerical
example, compactification scales of $1/R \sim 10^{-3}$ eV and $1/r
\sim 1$ TeV give $M \sim 100$ TeV.

The majority of the analyses on the collider phenomenology of
extra dimensions~\cite{collider} has been on the ADD scenario in
which the graviton propagates in the bulk, but where the SM fields
do not. Hence, the KK excitations of the graviton are the only
additional source of contributions to collider processes. While
the contributions of individual KK modes, with $4$D gravitational
strength, to collider processes is extremely small, the
compactification scale {$\mu$} is so small ($\mu \sim $ mm$^{-1}
\sim 10^{-3}$ eV) that a very large number of such modes
contribute in a TeV-scale collider process.  The net result can be
a significant deviation from the SM results.  Studies of various
collider processes typically give a bound on the string scale that
is the order of a TeV~\cite{collider}.

In comparison, the asymmetric scenario, in which SM fields, in
addition to the graviton, may propagate into one or more extra
dimensions of TeV$^{-1}$-size, will have a more direct effect in
high-energy collider processes.  Originating with the suggestion
by Antoniadis~\cite{Antoniadis}, some work has also been done for
the collider phenomenology of the scenario in which the SM gauge
bosons can propagate into the bulk, but where the SM fermions can
not~\cite{asymcoll}.  This includes the effects on EW precision
measurements~\cite{ew}, Drell-Yan processes in hadronic
colliders~\cite{muon}, $\mu^{+} \mu^{-}$ pair production in
electron-positron colliders~\cite{muon}, and recently multijet
production in very high-energy hadronic colliders~\cite{gstar}.
Typically, the bound on the compactification scale is $1$--$2$
TeV.  More recently, the case where all of the SM fields,
including the fermions, propagate into the extra dimensions has
been investigated~\cite{fermions,fermions2}.  These ``universal''
extra dimensions result in a much lower bound on the
compactification scale of a few hundred GeV.  The reason is that,
in this scenario, the tree-level couplings involving one KK mode
and two zero modes are not allowed.  Thus, the vertices involve a
pair of KK modes, and the KK modes need to be pair-produced, which
results in the suppression of the production cross sections.

In this work, we study the asymmetric compactification scenario
proposed in Ref.~\cite{asym}, in which only the SM gauge bosons
(and perhaps the Higgs boson) propagate into one of the
TeV$^{-1}$-size extra dimensions.\footnote{However, our results
apply to any compactified string model in which the SM gauge
bosons propagate into one such extra
dimension.}\addtocounter{footnote}{-1}  More specifically, we
investigate the effects that the KK excitations of the electroweak
(EW) gauge bosons have on various $e^{+}e^{-}$ collider processes.
We calculate the modifications to the SM cross sections which
arise from the direct production and exchanges of KK excitations
of the EW gauge bosons.  Included in our study are dijet
production, associated Higgs production, single-photon production,
and muon pair production and Bhabha scattering.  Although the
compactification scale must be quite small ($\lesssim 2$ TeV) for
a $6\%$ effect to be observed at the LEP$2$ energies, we find
substantial deviations from the SM cross sections for a future
collider with $\sqrt{s} = 600$ GeV (\textit{e.g.}, a
compactification scale of $3.5$ TeV produces an effect of $20\%$
for muon pair, dijet, and Higgs production), and an even greater
effect is predicted at higher energies. Our paper is organized as
follows. We begin by discussing our formalism in Section 2. In
Section 3, we consider charged lepton pair production, including
Bhabha scattering and muon pair production. We investigate dijet
production in Section 4.  In Section 5, we study Higgs production,
both for the case of SM as well as supersymmetric (SUSY) Higgs
doublets.  We discuss the production of a single photon with a
neutrino pair in Section 6. Section 7 contains our conclusions.

\vspace{0.5cm}

\noindent {\bf 2.  Formalism}

\vspace{0.2cm}

\noindent As it is unlikely that a significant number of
lowest-lying KK excitations will be directly produced in the near
future at $e^{+} e^{-}$ colliders, we focus here on tree-level
processes involving the exchanges of KK excitations of the EW
gauge bosons.  This necessarily restricts us to the case where the
initial and final state fermions are each confined to the SM D$_3$
brane since otherwise, as we shall see shortly, the couplings for
the processes of interest here are either zero or highly
suppressed when KK number is not conserved, depending on which
fields see the extra dimensions.  Thus, for simplicity, we
consider the case in which the EW gauge bosons can propagate into
a single extra TeV$^{-1}$-size dimension, but where the SM
fermions are restricted to lie on the SM wall (at the same
location in the extra dimension).  As for the Higgs boson, we also
restrict it to lie on the SM D$_3$ brane when we investigate
associated Higgs production (else there is either no effect or a
suppressed effect).

The terms in the $5$D Lagrangian density that involve fermion
fields contain a delta function to constrain the SM fields to the
usual $4$ spacetime dimensions, and similarly for terms involving
the Higgs field in the case in which the Higgs does not propagate
in the bulk.  The relevant parts of the $5$D Lagrangian density
can be expressed as

\vspace{-8pt} \begin{equation} \mathcal{L}_5  =
\mbox{\raisebox{-.6ex}{\huge $[$}} \mid \! \Xi_M H \! \mid^2 + i
g_{{}_5} \bar{f} \gamma^{\mu} D_{\mu} f \delta (y)
\mbox{\raisebox{-.6ex}{\huge $]$}}\, ,
\end{equation}

\noindent where $f$ is a SM fermion field, $H$ represents the
Higgs doublet(s), $D_M$ with $5$D spacetime index $M \in
{0,1,\ldots,4}$ is the $5$D generalization of the usual covariant
derivative $D_{\mu}$ with $4$D spacetime index $\mu$, the factor
$\mid \! \Xi_M H \! \mid^2$ denotes $\mid \! D_M H \! \mid^2$ for
a Higgs propagating in the bulk and $\mid \! D_{\mu} H \! \mid^2
\delta(y)$ for a Higgs localized to the SM boundary, and the
compactified extra dimension coordinate $y$ is related to the
radius of the extra dimension $r$ by $y  = r \phi$.  We consider
compactification on a $S^1 / Z_2$ orbifold with the identification
$\phi \rightarrow -\phi$.  In terms of the compactified dimension
$y$, a field $A_{\mu}(x,y)$ can then be Fourier expanded as

\vspace{-8pt} \begin{equation} A_{\mu} (x,y) = \frac{1}{\sqrt{\pi
r}}\mbox{\raisebox{-.6ex}{\huge $[$}}A_{\mu 0} (x) + \sqrt{2}
\sum_{n=1}^{\infty}A_{\mu n} (x)
\cos(n\phi)\mbox{\raisebox{-.6ex}{\huge $]$}} \, .
\end{equation}

\noindent The normalization for the $n = 0$ field $A_0 (x)$ is
one-half that of the $n > 0$ field $A_n (x)$.  Integration over
$y$ results in a tower of $A_{\mu n} (x)$ KK excitations.  The $n
= 0$ modes of the $5$D photon, $W^{\pm}$, and the $Z$ are
identified as the SM photon, $W^{\pm}$, and Z.  The $n > 0$ KK
modes of these fields are represented with a star ($\star$), as in
$\gamma_n^{\star}$.  As it is convenient to refer to the $n = 0$
and $n > 0$ modes separately, unless specified otherwise, $n$
shall henceforth imply strictly $n > 0$.

The detailed procedure for integrating over the fifth dimension
$y$ to obtain, in the effective $4$D theory, the factors for the
allowed vertices involving KK excitations of the EW gauge bosons
is similar to the procedure given in Ref.~\cite{gstar} for the
couplings of quarks and gluons to KK excitations of the gluons.
The difference is that any Higgs fields confined to the boundary
induce mixing terms~\cite{rescale} between the EW gauge bosons and
their KK excitations.  This causes a slight reduction in the
couplings involving KK excitations compared to the case in which
the Higgs fields propagate into the extra dimension; in addition,
previously forbidden couplings are allowed with a suppression
factor. However, because this mixing is highly suppressed (by a
factor of $\sim m_W^2 / \mu^2$), we only use the couplings given
by the case in which the Higgs fields propagate into the bulk, to
a good approximation. The couplings for the vertices involving KK
excitations of the $A^{\mu}$, $B^{\mu}$, and $C^{\mu}$ fields are
then given in terms of the corresponding couplings between SM
fields by the factors given in Ref.~\cite{gstar}; the
corresponding couplings for KK excitations of the photon,
$W^{\pm}$, and $Z$ are then related to the former via the usual
mixing relations.  In particular, there is a factor of $\sqrt{2}$
relative to the analogous SM coupling for a vertex involving a
single KK excitation and two SM fermions confined to the SM D$_3$
brane, which originates from the different rescaling of the $n =
0$ and $n  > 0$ modes necessary to obtain canonically normalized
kinetic energy terms in the effective $4$D Lagrangian
density~\cite{rescale}.  Also worth noting is that a single
$X_n^{\star}$ can not couple to two (or three) $Y$'s, where $X,Y
\in \{\gamma,W^{\pm},Z\}$ and $X_n^{\star}$ denotes a KK
excitation of gauge boson $X$ with mode $n > 0$.  In fact,
$X^{\star}$'s can only couple to other $X^{\star}$'s and
$Y^{\star}$'s if the modes
$n_{{}_{1}}$,$n_{{}_{2}}$,$\ldots$,$n_{{}_N}$ of these KK
excitations satisfy KK number conservation:

\vspace{-13pt} \begin{equation} \label{eq:modes} \mid \!
n_{{}_{1}} \, \pm \, n_{{}_{2}} \, \pm \, \cdots \, \pm \,
n_{{}_{N-1}} \! \mid \,= n_{{}_N} \, .
\end{equation}

Another difference between the Feynman rules for the EW gauge
bosons $\{X\}$ and their KK excitations $\{X_n^{\star}\}$ is that
the KK excitations are considerably heavier.  For a
compactification scale $\mu = 1/r$, the mass of the
n${}^{\textit{th}}$ KK excitation is:

\vspace{-13pt} \begin{equation} m_{X_n^{\star}} = \sqrt{m_X^2 +
n^2 \mu^2} \, .
\end{equation}

\noindent The $X_n^{\star}$ propagator is that of a usual massive
gauge boson:

\vspace{-8pt} \begin{equation} \label{eq:prop} - i
\Delta_{\mu\nu}(X_n^{\star},p^2) = -i \frac{g_{\mu \nu}  -
\frac{p_{\mu} p_{\nu}}{m_{X_n^{\star}}^2}}{p^2  -
m_{X_n^{\star}}^2  +  i m_{X_n^{\star}} \Gamma_{X_n^{\star}}} \, .
\end{equation}

\noindent  At tree-level, the $W_n^{\star}$ and $Z_n^{\star}$ have
the same decay rates as the $W$ and $Z$ except for a factor of $2
\frac{m_{W_n^{\star}}}{m_W}$ for the $W_n^{\star}$ and similarly
for the $Z_n^{\star}$; also, the KK excitations are heavy enough
to include the top quark in the decay rates.  The
$\gamma_n^{\star}$ decays to fermion pairs with total width
$\frac{14}{3}\alpha(m_{\gamma_n^{\star}})
m_{\gamma_n^{\star}}$.\footnote{We neglect the top quark mass
relative to the very heavy KK excitations.}
For diagrams where a $X_n^{\star}$ exchanges between two fermion
pairs (\textit{e.g.}, in $e^{+}e^{-} \rightarrow  t \bar{t} $),
there is the usual diagram with the $X$ propagator in addition to
a tower of diagrams with $X_n^{\star}$ propagators, or,
equivalently, an effective propagator given by the sum

\vspace{-8pt} \begin{equation} \Delta_{\mathit{eff}}(X,p^2)  =
c_{X_0}\Delta(X_0,p^2) \, +  \, \sum_{n=1}^{\infty}
c_{X_n^{\star}}\Delta(X_n,p^2) \, .
\end{equation}

\noindent The factors $\{c_{X_n}\}$ incorporate the different
$f_1$-$f_2$-$X$ and $f_1$-$f_2$-$X_n^{\star}$ vertex factors
(\textit{i.e.}, $c_{X_0}  = 1$, $c_{X_{n>0}}  =  2$).  This
effective propagator can be generalized to the case of arbitrary
vertices via adjustment of the $c_{X_n}$ factors (including
setting $c_{X_n}$ equal to zero when either vertex is forbidden).

\vspace{0.5cm}

\noindent {\bf 3.  Muon Pair Production and Bhabha Scattering}

\vspace{0.2cm}

\noindent Muon pair production is among the best prospects for the
indirect observation of KK excitations of the EW gauge bosons in
$e^{+}e^{-}$ processes because it can be measured with relatively
high precision at present and upcoming colliders while still
producing rather substantial deviations from the SM cross section
in comparison with other processes.  Muon pair production has
already been investigated elsewhere in the literature~\cite{muon}
for the purpose of setting present and future $e^{+}e^{-}$
collider bounds.  We primarily include this process here as a
standard by which to compare our other results for other
processes, but it also serves as a check on our calculations.  We
keep the first $100$ KK excitations in our analysis for purely
direct-channel processes and $25$ KK excitations for processes
with both direct- and cross-channel Feynman diagrams; in addition,
we consider $e^{+}e^{-}$ collider energies from LEP$2$ energies to
$1.5$ TeV and compactification scales up to $10$ TeV.

For charged lepton pair production, the KK excitations of the EW
gauge bosons manifest themselves through a tower of diagrams with
$\gamma_n^{\star}$ and $Z_n^{\star}$ propagators.  The net
tree-level effect of the $\gamma_n^{\star}$'s and $Z_n^{\star}$'s
on charged lepton pair production is the replacement of the SM
propagator by an effective KK propagator.  We employ gauge
invariance to drop the second term in Eq.~(\ref{eq:prop}) in our
analysis of fermion pair production. The effective moduli-squared
of the propagators for direct-channel $\gamma_n^{\star}$ and
$Z_n^{\star}$ exchange and the corresponding direct-channel
$\gamma_n^{\star}$-$Z_n^{\star}$ interference are
thus\footnote{When generalizing to the case where the EW gauge
bosons may propagate into more than one large extra dimension, the
sum in the effective propagator is formally divergent.  However,
various solutions to this problem have been proposed in the
literature~\cite{convergence}.}\addtocounter{footnote}{-2}

\vspace{-3pt} \begin{eqnarray} \label{eq:Dseff} \mid \!
D_{\mathit{eff}}(\gamma,s) \! \mid^2 \!\!\!&  = &\!\!\!
\frac{1}{2}\sum_{m,n=0}^{\infty} c_{\gamma_m} c_{\gamma_n}
\frac{s'_{\gamma_m}
  s'_{\gamma_n}
  +  m_{\gamma_m} \Gamma_{\gamma_m} m_{\gamma_n} \Gamma_{\gamma_n}}{(s_{\gamma_m}^{2}  +
m_{\gamma_m}^{2}\Gamma_{\gamma_m}^{2}) (s_{\gamma_n}^2  +
m_{\gamma_n}^2\Gamma_{\gamma_n}^2)} \nonumber\\
\mid \! D_{\mathit{eff}}(Z,s) \! \mid^2 \!\!\!&  = &\!\!\!
\frac{1}{2}\sum_{m,n=0}^{\infty} c_{Z_m} c_{Z_n} \frac{s'_{Z_m}
  s'_{Z_n}
  +  m_{Z_m} \Gamma_{Z_m} m_{Z_n} \Gamma_{Z_n}}{(s_{Z_m}^{2}  +
m_{Z_m}^{2} \Gamma_{Z_m}^{2}) (s_{Z_n}^2  +
m_{Z_n}^2 \Gamma_{Z_n}^2)} \\
\mbox{\raisebox{-.6ex}{\huge $[$}}D_{\mathit{eff}}^{\ast}(Z,s)
D_{\mathit{eff}}(\gamma,s) \!\!\! & + & \!\!\!
D_{\mathit{eff}}(\gamma,s)D_{\mathit{eff}}^{\ast}(Z,s)\mbox{\raisebox{-.6ex}{\huge
$]$}}\nonumber\\
\!\!\!&  = &\!\!\!  \sum_{m,n=0}^{\infty} c_{\gamma_m}
c_{\gamma_n} \frac{s'_{\gamma_m}
  s'_{Z_n}
  +  m_{\gamma_m} \Gamma_{\gamma_m} m_{Z_n} \Gamma_{Z_n}}{(s_{\gamma_m}^{2}  +
m_{\gamma_m}^{2}\Gamma_{\gamma_m}^{2}) (s_{Z_n}^2  +
m_{Z_n}^2\Gamma_{Z_n}^2)} \, . \nonumber
\end{eqnarray}

\noindent Here $s'_{X_n}$ is shorthand for the subtraction of
$m_{X_n}^2$ from $s$ (\textit{i.e.}, $s'_{X_n}  \equiv s -
m_{X_n}^2$) and $c_{X_n}$ represents the fact that the
$f$-$\bar{f}$-$X$ and the $f$-$\bar{f}$-$X_n^{\star}$ vertex
factors differ by a $\sqrt{2}$ (\textit{i.e.}, $c_{X_0}  =  1$,
$c_{X_{n>0}} \, = \, 2$).  (We make an exception in the effective
propagator equations by including the $n = 0$ and $n > 0$ modes
together for more compact notation.)  The effective propagator
formulae for cross-channel exchanges and interference are the same
as in Eq.~\ref{eq:Dseff} with the replacement of the
direct-channel Mandelstam variable $s$ by the cross-channel
variable $t$ (or $u$).  The interference between direct-channel
exchanges of $\gamma_n^{\star}$'s and cross-channel exchanges of
$Z_n^{\star}$'s is described by

\vspace{-8pt} \begin{eqnarray} \label{eq:Dteff}
\mbox{\raisebox{-.6ex}{\huge $[$}}D_{\mathit{eff}}^{\ast}(Z,t)
D_{\mathit{eff}}(\gamma,s) \!\!\!& + &\!\!\!
D_{\mathit{eff}}(\gamma,s)D_{\mathit{eff}}^{\ast}(Z,t)\mbox{\raisebox{-.6ex}{\huge
$]$}} \nonumber\\
\!\!\! & = & \!\!\! \sum_{m,n=0}^{\infty}
c_{\gamma_m} c_{\gamma_n} \frac{s'_{\gamma_m}
  t'_{Z_n}
  +  m_{\gamma_m} \Gamma_{\gamma_m} m_{Z_n} \Gamma_{Z_n}}{(s_{\gamma_m}^{2}  +
m_{\gamma_m}^{2}\Gamma_{\gamma_m}^{2}) (t_{Z_n}^2  +
m_{Z_n}^2\Gamma_{Z_n}^2)} \, .
\end{eqnarray}

\noindent  Finally, the effective propagator formulae for
direct-channel exchanges of $Z_n^{\star}$'s and cross-channel
exchanges of $\gamma_n^{\star}$'s are identical to
Eq.~\ref{eq:Dteff} with the replacement $\gamma \leftrightarrow
Z$.  For collider energies in the range $100$ GeV $\leq \sqrt{s}
\leq 1$ TeV and compactification scales $\mu > 1$ TeV, the sums in
the effective propagators converge quite rapidly.  Thus, they can
be truncated after only a few terms with negligible error.

\begin{figure}
\setlength{\abovecaptionskip}{0pt}
\centering{\includegraphics[bb=108 90 549 661.5]{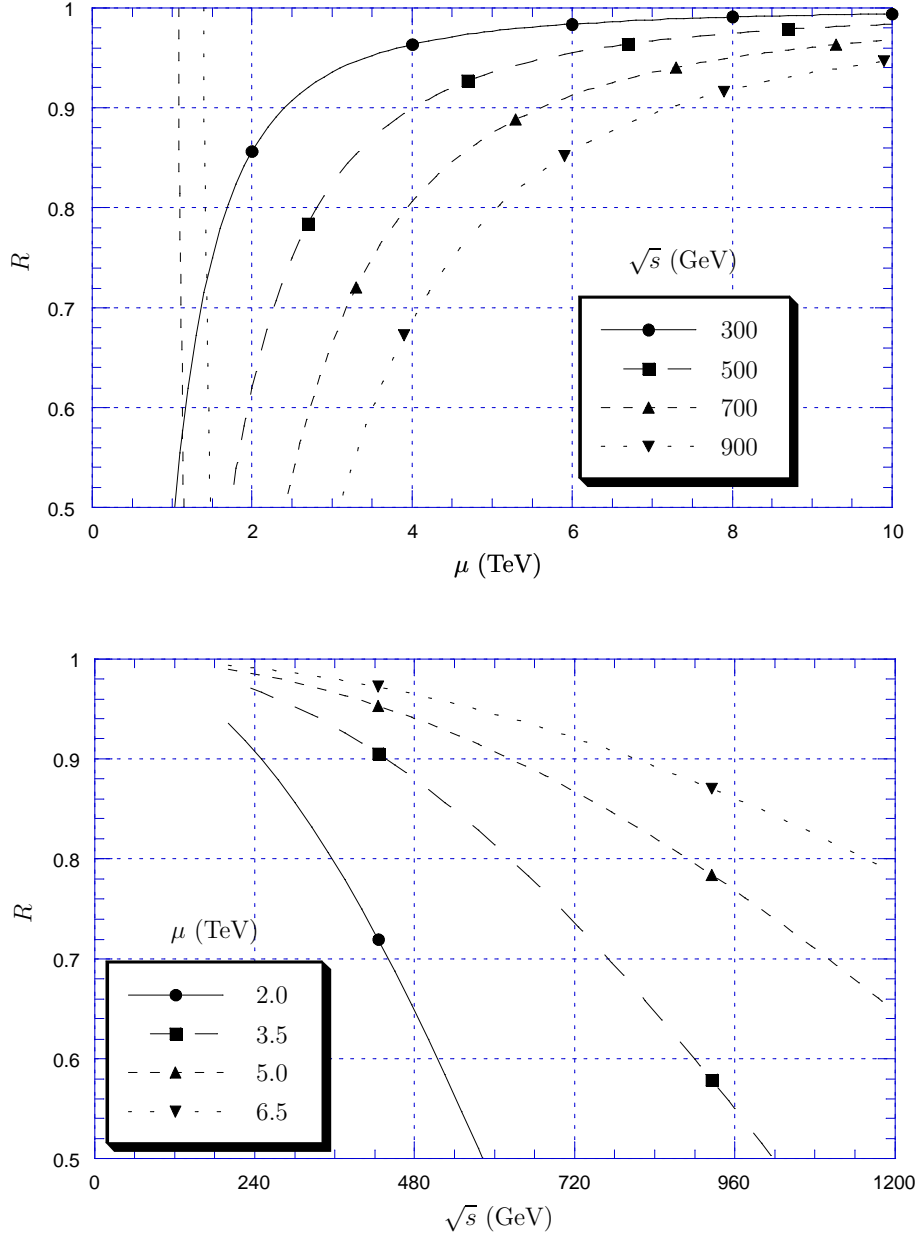}}
\vspace{-70pt} \caption{The contributions of the exchanges of
$\gamma_n^{\star}$'s and $Z_n^{\star}$'s to muon (or tau) pair
production are illustrated as a function of the compactification
scale $\mu$ for fixed values of the collider energy $\sqrt{s}$
(top), and as a function of $\sqrt{s}$ for specific choices of
$\mu$ (bottom).}\label{fig:muon}
\setlength{\abovecaptionskip}{0pt}
\end{figure}

The cross section for muon (or tau) pair production is easily
obtained via replacement of the usual SM propagator terms by the
effective propagator terms of Eq.~\ref{eq:Dseff}. Because the
compactification scale $\mu$ is a TeV or more and feasible values
of $\sqrt{s}$ for colliders in the present and not-too-distant
future range up to about a TeV, the primary effect of the KK
excitations of the EW gauge bosons arises from the interference of
the $n = 0$ (SM) mode terms with the $n > 0$ (KK) mode terms.
Since $(s - m_{X_{n>0}}^2) < 0$ for the ranges of $\sqrt{s}$ and
$\mu$ that we consider, the overall effect of the KK excitations
is a reduction in the muon (or tau) pair production cross section
as compared to the SM.  This effect is illustrated in
Fig.~\ref{fig:muon}, where the ratio

\vspace{-13pt} \begin{equation}\label{eq:Ratio} R \equiv
\frac{\sigma}{\sigma_{\mathit{SM}}} = \frac{\sigma_{\mathit{SM}} +
\sigma_{\mathit{KK}}}{\sigma_{\mathit{SM}}}
\end{equation}

\noindent is plotted for variation of compactification scale $\mu$
and collider energy $\sqrt{s}$ in the ranges $1$ TeV $\leq \mu
\leq 10$ TeV and $0.3$ TeV $\leq \sqrt{s} \leq 1.5$ TeV.  It is
clear that a very precise measurement of the cross section
($\lesssim 6\%$ uncertainty) is needed in order to observe a KK
effect at a LEP$2$ collider running at $\sqrt{s} = 200$ GeV for
$\mu \geq 2$ TeV. However, the effect increases significantly for
larger collider energies:  For example, a compactification scale
of $4$ TeV produces an effect of only a few percent at LEP$2$
energies, while it reduces the cross section by more than $30 \%$
at a TeV-scale $e^{+}e^{-}$ high-energy collider.

\begin{figure}
\setlength{\abovecaptionskip}{0pt}
\centering{\includegraphics[bb=108 90 549 661.5]{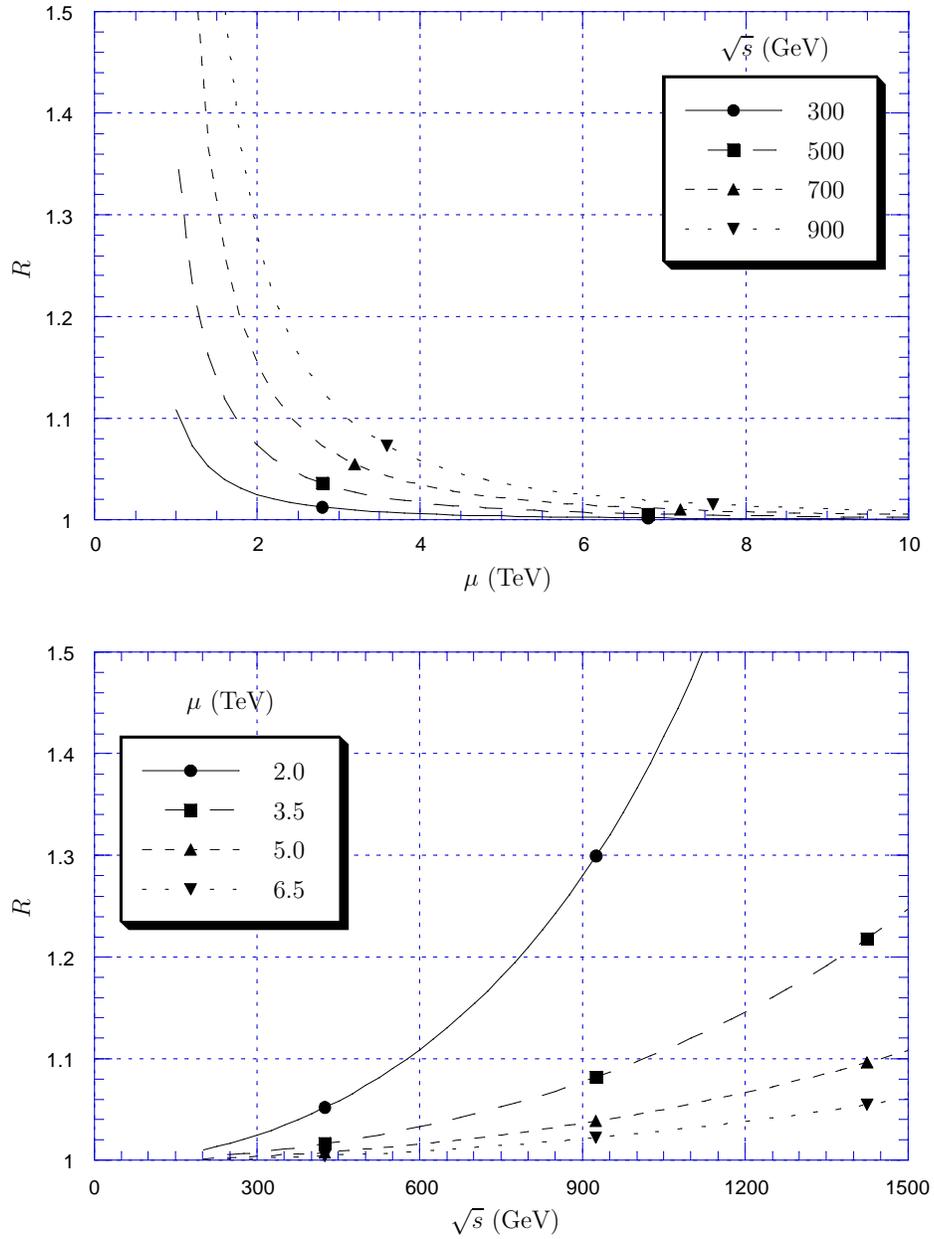}}
\vspace{-70pt} \caption{The same as Fig.~\ref{fig:muon}, but for
Bhabha scattering.}\label{fig:Bhabha}
\setlength{\abovecaptionskip}{0pt}
\end{figure}

Bhabha scattering involves the cross-channel exchanges of the
$\gamma$ and $Z$ as well as the direct-channel exchanges of muon
pair production.  Again the primary effect can be attributed to
the interference of the $n = 0$ mode with the $n > 0$ modes.
However, although this interference causes a reduction in the
direct-channel cross section (\textit{i.e.}, muon pair
production), it has the opposite effect in the cross-channel.
These competing effects lead to a smaller overall effect of the KK
excitations on Bhabha scattering as compared to muon pair
production.  This overall enhancement of the SM cross section is
depicted in Fig.~\ref{fig:Bhabha} for the same range of parameters
as in the muon case.
The enlargement of the SM cross section is less than $10 \%$ for
$\mu > 3$ TeV or $\sqrt{s} < 900$ GeV.  This makes Bhabha
scattering considerably less attractive for observable effects of
the KK states as compared to muon production.

\vspace{0.5cm}

\noindent {\bf 4.  Dijet Production}

\vspace{0.2cm}

\begin{figure}
\setlength{\abovecaptionskip}{0pt}
\centering{\includegraphics[bb=108 90 549 661.5]{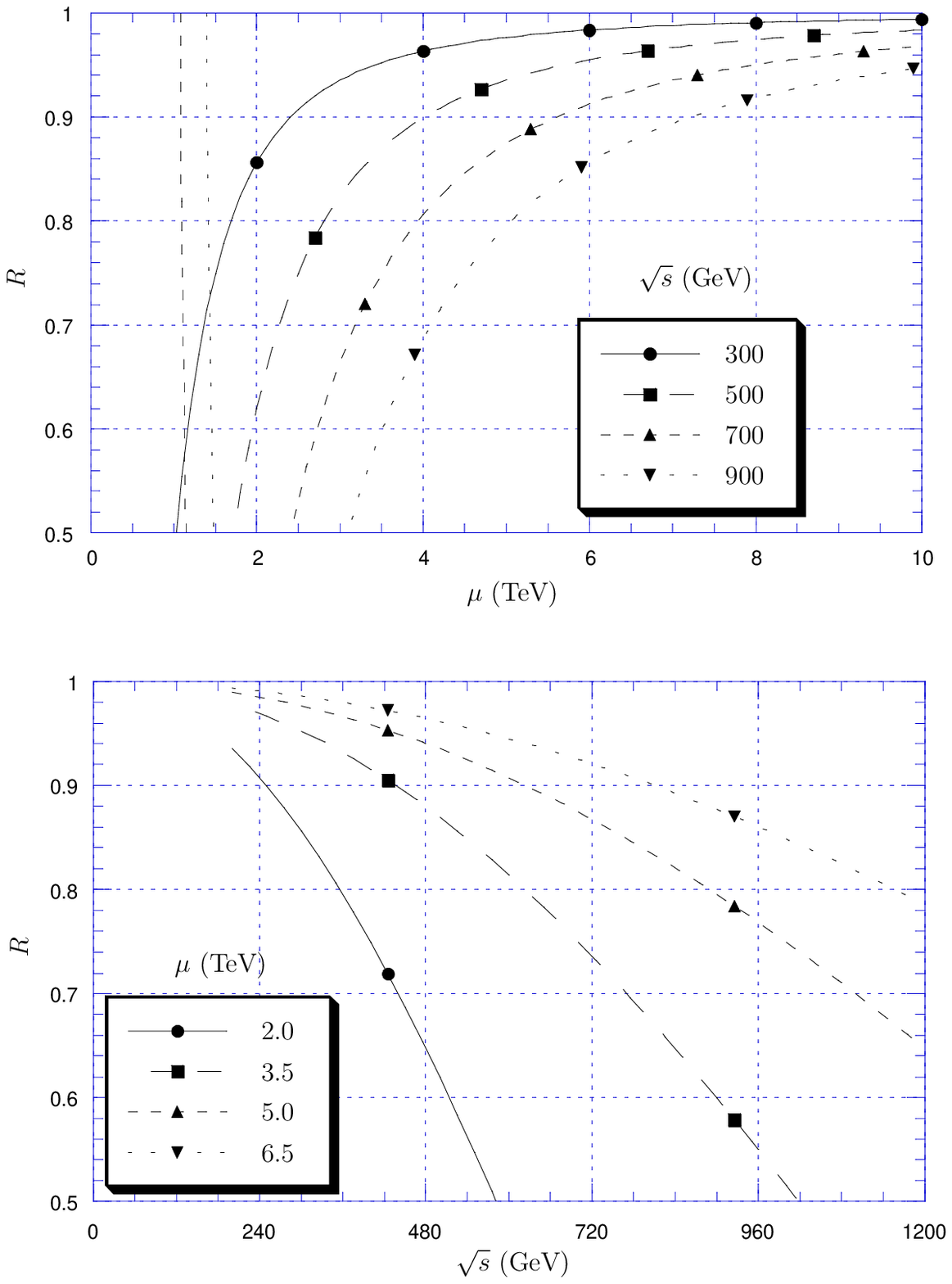}}
\vspace{-70pt} \caption{The same as Fig.~\ref{fig:muon}, but for
dijet production.}\label{fig:dijet}
\setlength{\abovecaptionskip}{0pt}
\end{figure}

\noindent The full dijet final state cross section is given by
summing $e^{+}e^{-} \rightarrow q \bar{q}$ over all quark flavors
for which $s > 4 m_q^2$.  Thus, top quark production is only
included for $\sqrt{s} > 350$ GeV.  We include the standard top
quark corrections in the cross section formulae, since the top
quark mass is significant compared to $\sqrt{s}$ for LEP energies,
but note that the top quark mass is negligible in comparison to
TeV-scale KK excitations.  The KK excitations result in the same
effective $s$-channel propagator expressions as in the case of
muon pair production (Eq.~\ref{eq:Dseff}).  As a result, the ratio
$R$ is virtually identical in the two cases.  The KK effect on
dijet final state production is plotted in Fig.~\ref{fig:dijet}.
As for muon production, a compactification scale of $3.5$ TeV
results in a reduction by $50\%$ at a TeV-scale collider, by
$12\%$ for $\sqrt{s} = 500$ GeV.

\vspace{0.5cm}

\noindent {\bf 5.  Higgs Production}

\vspace{0.2cm}

\noindent First, we consider the associated SM production of the
Higgs boson:  $e^{+}e^{-} \rightarrow Z H$.  Here, we take the $Z$
boson to be produced on-shell.  As discussed previously, the KK
contribution is either zero or strongly suppressed due to KK
number non-conservation unless the Higgs boson is confined to the
SM three-brane. The $Z$-$Z_n^{\star}$-$H$ coupling is non-zero in
this situation because the corresponding term in the $5$D
Lagrangian density contains a delta function to constrain the
Higgs boson to the SM wall. Therefore, we focus on this scenario
here.  The effect of the KK excitations of the $Z$ is described by
the replacement of the SM direct-channel $Z$ boson propagator by
the effective propagator given in Eq.~\ref{eq:Dseff}.  Although
the Higgs mass is important in limiting the available collider
energy range and determining the numerical value of the SM cross
section, it does not affect the ratio of the total cross section
given by the sum of the SM and KK contributions to the SM cross
section. In fact, since there is only one diagram involved, the
ratio $R$ is given by the effective propagator from
Eq.~\ref{eq:Dseff}:

\vspace{-0.3pt} \begin{equation} R = \mid \!
D_{\mathit{eff}}^{\star}(Z,s) \! \mid^2
  =  \frac{1}{2}\sum_{m,n=0}^{\infty} c_{Z_m} c_{Z_n} \frac{s'_{Z_m}
  s'_{Z_n}
  +  m_{Z_m} \Gamma_{Z_m} m_{Z_n} \Gamma_{Z_n}}{(s_{Z_m}^{2}  +
m_{Z_m}^{2} \Gamma_{Z_m}^{2}) (s_{Z_n}^2  + m_{Z_n}^2
\Gamma_{Z_n}^2)}
\end{equation}

\noindent Thus, not only is the ratio $R$ independent of the Higgs
mass, it is also independent of the Higgs model (to the point
where there are only exchanges of $Z$ or $Z_n^{\star}$ bosons).
For example, the ratio $R$ is the same in the SUSY Higgs doublet
case of $e^{+}e^{-} \rightarrow A h$ as in the SM case of
$e^{+}e^{-} \rightarrow Z H$. However, the total cross section
depends on the Higgs model, and this plays a strong role in
determining if the total cross section is significant enough to
observe the Higgs boson(s) (and if so, if it is large enough to
see a KK effect).

\begin{figure}
\setlength{\abovecaptionskip}{0pt}
\centering{\includegraphics[bb=108 90 549 661.5]{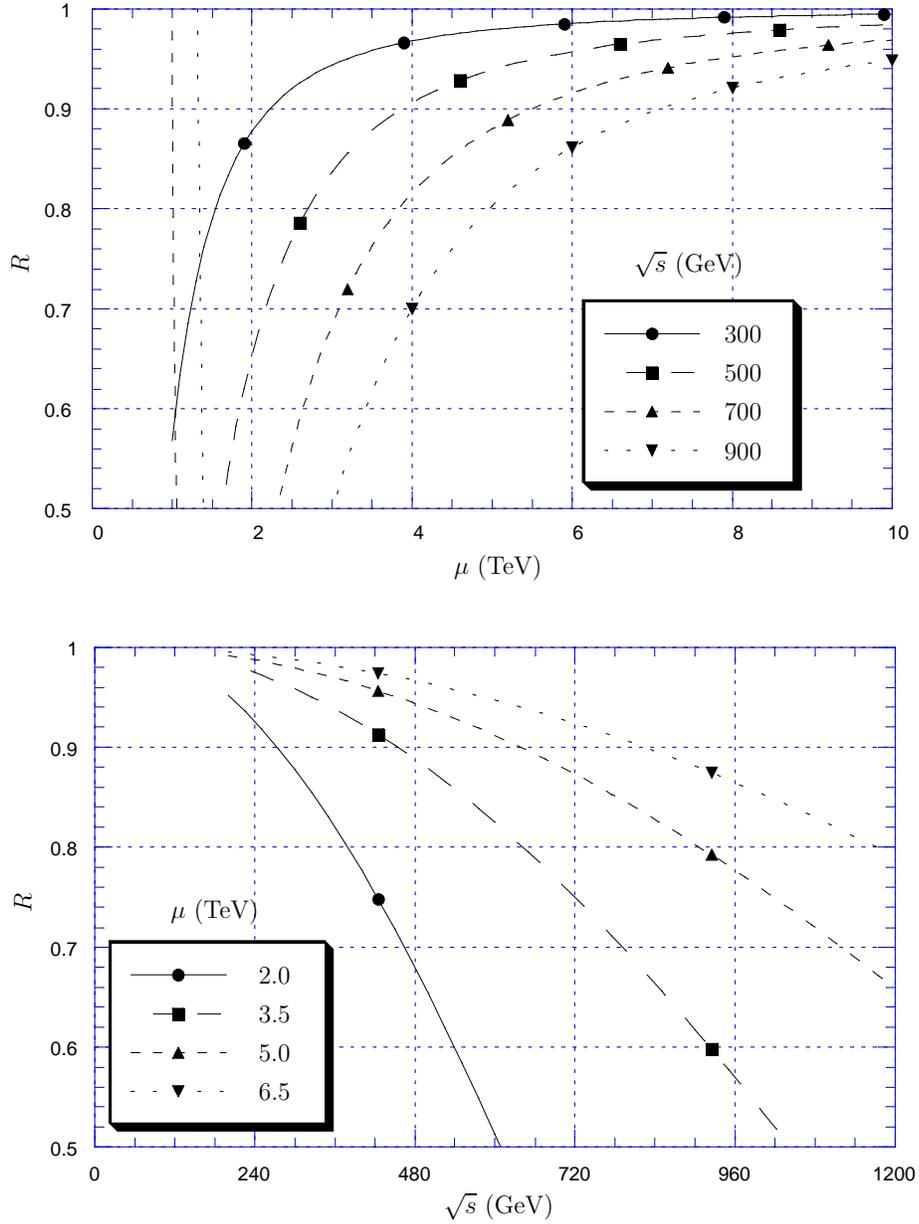}}
\vspace{-70pt} \caption{The contributions of the exchanges of
$Z_n^{\star}$'s to Higgs production are illustrated as a function
of the compactification scale $\mu$ for fixed values of the
collider energy $\sqrt{s}$ (top), and as a function of $\sqrt{s}$
for specific choices of $\mu$ (bottom).}\label{fig:Higgs}
\setlength{\abovecaptionskip}{0pt}
\end{figure}

The KK effect on Higgs production for processes (such as the SM
and SUSY Higgs doublet cases discussed above) in which there are
only exchanges of $Z$ or $Z_n^{\star}$ bosons is shown in
Fig.~\ref{fig:Higgs}, where $R$ is graphed as a function of the
compactification scale $\mu$ and collider energy $\sqrt{s}$ for
the same range of parameters as in the case of muon pair
production.
Although there are no photon exchanges in the Higgs case, the
effect of KK excitations of the $Z$ boson on Higgs production is
almost identical to the KK effect on muon pair production.
Lowest-lying KK excitations of the $Z$ boson with masses of about
$5$ TeV cause a $20\%$ reduction compared to the SM cross section
for a collider energy of $1$ TeV, whereas the reduction is only
$5\%$ at $\sqrt{s} = 500$ GeV, and only $2\%$ at the LEP$2$
energies. However, a compactification scale of $3.5$ TeV produces
at least a $10\%$ effect at collider energies beginning at $400$
GeV; the reduction is about half for a $1$ TeV energy collider.
This reduction in the overall cross section as compared to the SM
cross section also has a significant effect on the Higgs mass
bound, and this is Higgs model-dependent.

\vspace{0.5cm}

\noindent {\bf 6.  Neutrino Pair and Single Photon Production}

\vspace{0.2cm}

\noindent We first consider the case of neutrino pair production,
$e^{+}e^{-} \rightarrow \sum_{\ell} \nu_{\ell} \bar{\nu}_{\ell}$.
The production of muon or tau neutrino pairs only consists of
direct-channel $Z$ and $Z_n^{\star}$ exchanges described by the
effective propagator modulus-squared of Eq.~\ref{eq:Dseff},
whereas the production of electron neutrino pairs also includes
the cross-channel exchanges of $W$'s and $W_n^{\star}$'s.  The
modulus-squared of the effective propagator for this $t$-channel
production involving $W$'s and $W_n^{\star}$'s is the same as the
$s$-channel production involving $Z$'s and $Z_n^{\star}$'s with
the replacements $Z \rightarrow W$ and $s \rightarrow t$.
Similarly, the $s$-$t$ interference is given by Eq.~\ref{eq:Dteff}
with the replacement $\gamma \rightarrow W$.

\begin{figure}
\setlength{\abovecaptionskip}{0pt}
\centering{\includegraphics[bb=108 90 549 661.5]{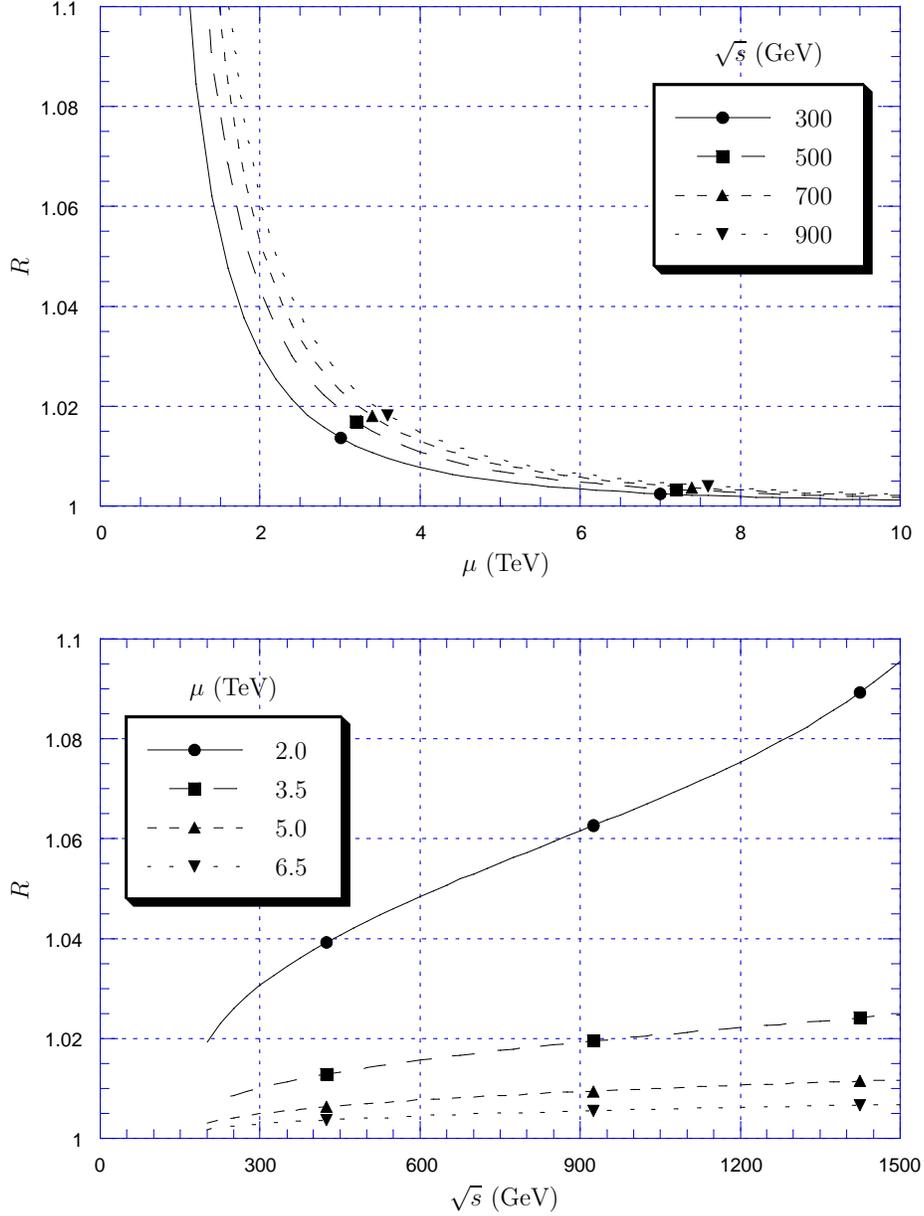}}
\vspace{-70pt} \caption{The contributions of the exchanges of
$W_n^{\star}$'s and $Z_n^{\star}$'s to neutrino pair production
are illustrated as a function of the compactification scale $\mu$
for fixed values of the collider energy $\sqrt{s}$ (top), and as a
function of $\sqrt{s}$ for specific choices of $\mu$
(bottom).}\label{fig:neutrino} \setlength{\abovecaptionskip}{0pt}
\end{figure}

As in the case of Bhabha scattering, the effect of the
direct-channel exchanges of the $Z$'s and $Z_n^{\star}$'s to
reduce the cross section and the competing effect of the
cross-channel exchanges of the $W$'s and $W_n^{\star}$'s to
increase the cross section as compared to the SM results in a
considerably smaller effect than processes such as muon pair
production and dijet production where there are only $s$-channel
exchanges of the EW gauge bosons.  Although the $s$-channel KK
effect to increase the cross section is larger than the
$t$-channel KK counter-effect percentage-wise, the SM $t$-channel
is dominant for neutrino pair production, which causes a slight
increase in the cross section as compared to the SM.  This is
illustrated in Fig.~\ref{fig:neutrino}, where the same ranges of
the collider energy $\sqrt{s}$ and compactification scale $\mu$
are employed as in the case of muon pair production.
The KK effect is smaller for neutrino pair production than for
Bhabha scattering; also, there appears to be far less dependence
on the variation of the collider energy.  The result is a KK
effect of less than $7\%$ even for a collider energy as high as a
TeV.

Single-photon production via $e^{+}e^{-} \rightarrow \nu \bar{\nu}
\gamma$ is somewhat more complicated.  Single-photon production
was considered in Ref.~\cite{photon} in the context of
$Z^{\prime}$ physics, where only the lowest-lying KK excitations
of the $W$ and $Z$ bosons were included.  Here, we extend their
results to include the $25$ lowest-lying states, but concede that
the effect depends almost exclusively on the first few states, and
primarily on the first.  The diagrams for single-photon production
are the same as for neutrino pair production with a photon
radiating off the incoming electron or positron or the internal
$W$ or $W_n^{\star}$.  The effect of the KK excitations results in
the same direct-channel effective propagator as the neutrino
production case, and the same cross-channel effective propagator
when the photon radiates off the incoming electron or positron.
However, for the case where the photon radiates off the internal
$W$ or $W_n^{\star}$, a difference arises from the coupling of the
photon to $W$'s and $W_n^{\star}$'s.  The
$\gamma$-$W$-$W_n^{\star}$ and $\gamma$-$W_m^{\star}$-$W_{n\neq
m}^{\star}$ couplings are forbidden due to KK number
non-conservation, as discussed in Section $2$.  On the other hand,
the diagram with the $\gamma$-$W_n^{\star}$-$W_n^{\star}$ coupling
has two propagators with KK excitations of the $W$ boson, which
are quite massive (TeV-scale). This suppresses the KK contribution
from this diagram in comparison to the contributions of the
diagrams in which the photon radiates off either of the incoming
particles.

\begin{figure}
\setlength{\abovecaptionskip}{0pt}
\centering{\includegraphics[bb=108 90 549 661.5]{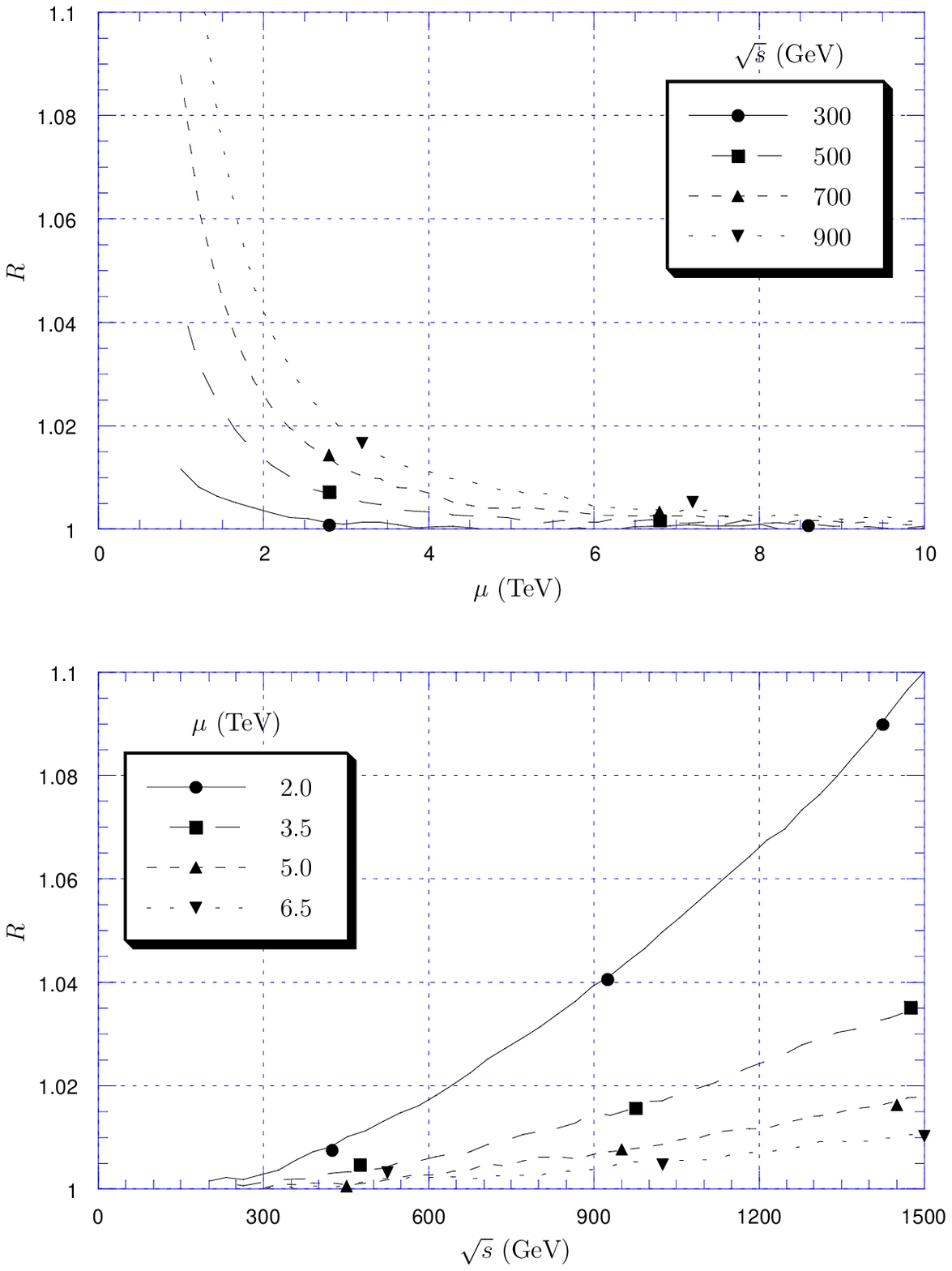}}
\vspace{-70pt} \caption{The same as Fig.~\ref{fig:neutrino}, but
for single-photon production.}\label{fig:photon}
\setlength{\abovecaptionskip}{0pt}
\end{figure}

The overall KK effect on single-photon production is very much
similar to the KK effect on neutrino pair production.
Graphically, the total KK contribution is shown in
Fig.~\ref{fig:photon}.
Again, the enlargement of the SM cross section is very small.  The
single-photon and neutrino pair production KK effects are almost
identical for large collider energies $\sim 1$ TeV, but the
single-photon case is more dependent on the collider energy,
resulting in an even smaller effect for LEP energies than the
neutrino pair case.  Again, the compactification scale must be
quite small $\lesssim 2$ TeV in order to see even a $5\%$ effect
for a collider with very high energy (TeV-scale).

\vspace{0.5cm}

\pagebreak[4] \noindent {\bf 7.  Conclusions}

\vspace{0.2cm}

\noindent We have investigated the phenomenology of the KK
excitations of the EW gauge bosons for a class of string-inspired
models in which the SM gauge bosons propagate into one TeV-scale
compact extra dimension, but where the SM particles are confined
to the usual SM three-brane.  Specifically, we have examined the
effects that these KK excitations have on the cross sections for
various processes at present and future high energy $e^{+}e^{-}$
colliders.  Included in our study were Bhabha scattering and muon
pair production, dijet production, Higgs production, and neutrino
pair and single-photon production.  Exclusively direct-channel
processes, namely, muon, dijet, and Higgs production, produced a
considerably greater effect than processes with both direct- and
cross-channel Feynman diagrams, \textit{i.e.} Bhabha scattering
and neutrino and single-photon production.  This is due to the
competing effects of the effective propagators for $s$-channel
exchanges and $t$-channel exchanges:  The primary effect of the KK
excitations arises from the interference of the $n = 0$ (SM) mode
exchanges with the $n > 0$ (KK) mode exchanges, which results in a
reduction of the modulus-squared of the effective propagator and
thus the corresponding amplitude-squared for direct-channel
exchanges, and an opposing enlargement for cross-channel exchanges
and the $s$-$t$ interference.

The KK excitations of the EW gauge bosons would be particularly
elusive for detection at LEP$2$ energies, where the largest effect
for the processes that we examined is below $6\%$ for even small
compactification scales such as $\sim 2$ TeV and below $3\%$ for
compactification scales of $\sim 3.5$ TeV. Thus, quite precise
measurements as well as very low compactification scales would be
necessary for hints of KK excitations of the EW gauge bosons at
LEP energies.  However, the effects are considerably greater for
prospective high energy colliders.  For example, a $500$ GeV
collider can see about a $20\%$ reduction in the cross section for
muon pair production compared to the SM if the lowest-lying KK
excitations have masses of $\sim 3$ TeV, and a $10\%$ reduction
for KK masses starting at $\sim 4$ TeV.  A very high energy (TeV)
collider could probe compactification scales up to $5$ TeV and
find a $20\%$ effect, and up to $7$ TeV with a $10\%$ effect;
meanwhile, a smaller compactification scale of $3$ TeV reduces the
cross section by half compared to the SM background.

We found that the KK excitations of the EW gauge bosons could play
an important role on the discovery of the Higgs by enhancing the
Higgs production cross section significantly. This is true for a
Higgs boson that is confined to the SM three-brane, else the
coupling of the Higgs to a single KK excitation of a gauge boson
is zero.

Here we address the differences between KK excitations and other
new physics that might produce a similar collider signal. In
particular, $W^{\prime}$ and $Z^{\prime}$ physics produce the same
effects as the lowest-lying KK excitations of the $W$ and the $Z$,
except that the couplings of the $W^{\prime}$ and $Z^{\prime}$ to
fermions can be different, and there are no restrictions on how
many $W^{\prime}$'s and/or $Z^{\prime}$'s can couple to SM gauge
bosons. Although the KK case involves an infinite tower of
$W_n^{\star}$'s and $Z_n^{\star}$'s, the primary effect arises
from the interference between the $n = 0$ (SM) and $n = 1$ (KK)
modes, which is exactly the effect of the $W^{\prime}$ and
$Z^{\prime}$. In the case of multiple $Z^{\prime}$'s, for example,
if the various $Z^{\prime}$'s have masses that are not integral
multiples of the smallest $Z^{\prime}$ mass, then this would
clearly be different from the KK tower formed by a SM $Z$ boson
that propagates into one extra dimension. Also, there has been
abundant interest in $Z^{\prime}$ models with restricted couplings
to fermions, such as leptophobic $Z^{\prime}$'s that couple to
quarks but not to leptons, which seek to explain discrepancies
between SM theoretical predictions and experimental measurements
for particular processes without destroying fine agreement with
processes such as charged lepton production at the $Z$ pole.  In
these cases, the couplings are different from the KK model that we
consider here, where the KK excitations couple to all fermions
with a $\sqrt{2}$ relative to the SM couplings. However, it is
also possible to construct models in which some fermions see extra
dimensions while others do not. For example, if the leptons see an
extra dimension while the quarks are confined to the usual SM
wall, then the situation will mimic the behavior of leptophobic
$Z^{\prime}$ physics for $e^{+}e^{-}$ processes.\footnote{For this
KK case, the $e q \rightarrow Z_n^{\star} \rightarrow e q$
cross-channel process does not vanish, whereas a $Z^{\prime}$ can
not couple to the leptons.} Finally, we considered a model in
which all of the EW gauge bosons propagate into the same extra
dimension.  In this case, if there is a $Z_1^{\star}$ with a mass
of $3$ TeV, then there are also $\gamma_1^{\star}$'s and
$W_1^{\star}$'s with masses that are approximately $3$ TeV as
well.  However, it is also possible that the various EW gauge
bosons propagate into different extra dimensions with different
compactification scales, or that some do not see extra dimensions
at all.  All in all, there is are several differences between the
KK and $Z^{\prime}$ effects that can be calculated for various
processes, but the general behavior is quite similar. The chief
test would come from very high energy colliders.  If a
$Z^{\prime}$ or $Z^{\star}$ is detected, then a search at twice
that scale that fails to find a $Z_n^{\star}$ would clearly reveal
that it is a $Z^{\prime}$ and not a $Z^{\star}$, and a search that
finds a $Z_2^{\star}$ at the correct scale only leaves a small
probability that this is coincidentally a second $Z^{\prime}$ with
twice the mass of the first (in which case a search at three times
the first scale could reduce this probability even further or
decide in favor of the $Z^{\prime}$).

\vspace{0.5cm}


\vspace{0.2cm}

\noindent We are grateful to K.S. Babu, D.A. Dicus, J.D. Lykken,
and J.D. Wells for useful discussions.  This work was supported in
part by the U.S. Department of Energy Grant Numbers
DE-FG03-98ER41076 and DE-FG02-01ER45684.

\vspace{0.5cm}

\pagebreak[4]
\bibliographystyle{unsrt}

\end{document}